# Allotropic transition of Dirac semimetal α-Sn to superconductor β-Sn induced by irradiation of focused ion beam

Kohdai Inagaki[1,*], Keita Ishihara[1,*], Tomoki Hotta[1], Yuichi Seki[1], Takahito Takeda[1], Tatsuhiro Ishida[2], Daiki Ootsuki[2], Ikuto Kawasaki[3], Shin-ichi Fujimori[3], Masaaki Tanaka[1,4], Le Duc Anh[1,4,5,†], and Masaki Kobayashi[1, 4,‡]

[1]*Department of Electrical Engineering and Information Systems, The University of Tokyo, 7-3-1 Hongo, Bunkyo-ku, Tokyo 113-8656, Japan*

[2]*Graduate School of Human and Environmental Studies, Kyoto University, Kyoto 606-8501, Japan*

[3]*Materials Sciences Research Center, Japan Atomic Energy Agency, Sayo-gun, Hyogo 679-5148, Japan*

[4]*Center for Spintronics Research Network, The University of Tokyo, 7-3-1 Hongo, Bunkyo-ku, Tokyo 113-8656, Japan*

[5]*PRESTO, Japan Science and Technology Agency, 4-1-8 Honcho, Kawaguchi, Saitama 332-0012, Japan*

*These authors contributed equally to this work.
†Corresponding author: anh@crys.t.u-tokyo.ac.jp
‡Corresponding author: masaki.kobayashi@ee.t.u-tokyo.ac.jp

## Abstract

Diamond-type structure allotrope α-Sn is attracting much attention as a topological Dirac semimetal (TDS). In this study, we demonstrate that α-Sn undergoes a phase transition to another allotrope β-Sn with superconductivity at low temperature by irradiating with a focused Ga ion beam (FIB). To clarify the transition mechanism, we performed X-ray photoemission spectroscopy (XPS) measurements on an α-Sn thin film irradiated with FIB and an as-grown α-Sn thin film. The XPS results suggest that the local annealing, which is one of the side effects of FIB, causes the transformation from α-Sn into β-Sn. Furthermore, the difference in the chemical states between α-Sn and β-Sn can be quantitatively explained by the crystal structures rather than the degree of metallicity reflecting the conductivity. These results propose a new way of fabricating TDS/superconductor in-plane heterostructures based on α-Sn and β-Sn.

**Introduction**

Topological materials (TMs), such as topological insulators (TIs) [1,2], topological Dirac semimetals (TDSs) [3,4,5], and topological superconductors (TSCs) [6,7], have been actively studied owing to their interesting physical properties originated from the topology of the band structures. These TMs host linear dispersions in the electronic structures, called Dirac cones, either at the surface or in the bulk band structure. Because the electrons at Dirac cones possess high mobility and light effective mass with robust spin-momentum locking, TMs have shown great potential for device applications such as spin-orbit torque magnetic random-access memory, high-performance field effect transistors (FET) and quantum computers [6–12]. Among these TMs, TDSs attract much attention as a parent phase of many other topological phases. They also possess important potential for applications due to a variety of physical properties derived from the three-dimensional (3D) Dirac cones in the bulk, such as ultra-high mobility [13,14], light cyclotron mass [15], giant linear longitudinal magnetoresistance [13,15], Fermi arcs [16,17], and chiral anomaly [18–21].

Among the known TDSs such as $Na_3Bi$ and $Cd_3As_2$ [4,14], α-Sn has recently attracted considerable attention as a promising material candidate with a rich topological phase diagram, environment-friendly, and compatible with semiconductor technology [22-25]. In α-Sn, a band inversion between the *s* and *p* bands occurs [22] due to strong spin-orbit interaction, while the inverted light hole (LH) band and the heavy hole (HH) band degenerate at the Γ point. Under compressive in-plane strain, as in the case when α-Sn is grown on InSb, the LH and HH bands cross at two points along the $k_z$ axis, which are two Dirac cones, and the α-Sn thin film is transformed to the TDS phase [23–25]. Recently, Anh *et al*. have succeeded in the epitaxial growth of high-quality TDS α-Sn on an InSb (001) substrate and observed the high quantum mobilities due to the topological states [26]. On the other hand, another allotrope of Sn, β-Sn, is a metal with a tetragonal crystal structure and exhibits BCS superconductivity at low temperatures (<4 K). Interestingly, it is known that semimetallic α-Sn is transformed to metallic β-Sn upon heating, which provides an avenue to incorporate superconductivity into the already rich topological phase of α-Sn [27].

In this paper, we studied the transport and electronic structure properties of an α-Sn thin film being irradiated by a focused Ga ion beam (FIB) and found that the film exhibits superconductivity below 4 K. Transmission electron microscopy (TEM) and X-ray photoemission spectroscopy (XPS) are performed to clarify the transition mechanism and the electronic states of the α-Sn irradiated with FIB. The observations suggest that the FIB treatment alters α-Sn to β-Sn superconductor. Based on the findings, we discuss the effects of FIB and the details of the chemical states. Generally, FIB is used as a processing method to pattern device structures, prepare samples for TEM, and fabricate subtractive 3D structures in the nm scale by irradiating focused and accelerated ion beams [28–30]. In addition to the primary sputtering process, FIB leads to side effects on the irradiated area: ion implantation, surface amorphization, physical damage, and the local annealing [31,32]. These additional processes will affect the chemical states and/or the crystal structures.

**Experiments**

An α-Sn film with a thickness of 70 nm was epitaxially grown on an InSb (001) substrate without any surfactant by a molecular beam epitaxy (MBE) system equipped with effusion cells of Sn and III–V elements. The growth conditions are the same as the previous study [26], and growth of α-Sn on InSb causes in-plane compressive strain which makes α-Sn become TDS. Then, the α-Sn sample was cleaved into a chip shape, and a half area of the chip was irradiated with Ga FIB by a V400ACE FIB system. We used FIB with an acceleration voltage of 30 kV, a current value of 7.7 pA, and an irradiation pitch of 5 nm. (Fig. 1(a)). The XPS measurements were conducted in the helical undulator beamline BL23SU of SPring-8. The incident photon energy ($h\nu$) was 1400 eV, where the inelastic mean free path is about 2.0 nm [33]. The measurements were conducted with an SES-2002 photoelectron analyzer under an ultrahigh vacuum below $1.0 \times 10^{-8}$ Pa at room temperature (RT). The monochromator resolution $E/\Delta E$ was about 10,000 and the beam spot size was $200 \times 100$ μm$^2$ [34]. The total energy resolution was about 0.6 eV. The binding energies were also calibrated by measuring the Fermi level ($E_F$) of evaporated gold, which was electrically contacted to the sample.

**Results and Discussion**

Figure 1(b) shows the temperature dependence of the resistance of the α-Sn thin film irradiated with FIB and the superconducting transition below 4 K is observed. Either Sn-In alloy [35] or β-Sn [36] is a possible origin of the superconducting behavior in the irradiated area, in which Sn-In alloy is possibly formed by the interdiffusion of In into α-Sn from the substrate InSb. In addition, an anomalous sharp peak in resistance is observed just before the superconducting transition temperature around 3.8K in Figure 1(b). This behavior has also been observed in the α-Sn/β-Sn heterostructures in previous study [27] and can be explained by a suppression of the DOS near the Fermi level when Cooper pairs are formed in the normal conducting α-Sn by the superconducting proximity effect from the FIB irradiated α-Sn region.

To investigate the crystalline structure of the FIB-irradiated α-Sn area, we use a cross-sectional TEM image of the sample. Figure 2(a) is a cross-sectional TEM lattice image of the as-grown α-Sn region, which shows epitaxial growth on the InSb buffer layer retaining the diamond-type crystal structure in the entire film. Figure 2(b) shows a cross-sectional TEM lattice image of the FIB-irradiated α-Sn area. The morphology of Sn is largely changed by the FIB irradiation, and multiple crystal grains are formed. These results indicate that the FIB-irradiated superconducting regions have a polycrystalline structure, which was formed from the diamond-type crystal structure of α-Sn due to direct ion collisions and local annealing.

Figure 3(a) shows the wide XPS spectra of the as-grown and the irradiated areas in the α-Sn/InSb thin film. Here, the spectra of the as-grown and the irradiated areas are normalized to the background height at around a binding energy ($E_B$) of 1200 eV. In the wide spectra, peaks derived from Sn and O elements are mainly observed. The Sn peak is from the Sn constituent components of the thin film. The O peak likely originates from surface oxidization and/or extrinsic polar molecules in the air such as $H_2O$ and CO. Since the signal from In ($E_B \sim 444$ eV) was not observed, the existence of In ions is negligible near the surface of the film. This result suggests that the irradiated area does not become a superconducting Sn-In alloy. A signal from Ga ($E_B \sim 1116$ eV) in the irradiated area is too small to observe even if it exits, indicating that the effect of the Ga ion implantation caused by FIB near the surface is negligible.

Figure 3(b) shows the Sn $3d$ spectra of the as-grown film and the irradiated area. The spectra are normalized to the maximum intensity. While the Sn $3d_{5/2}$ peaks at 486.50 eV correspond to oxidized $Sn^{2+}$ (SnO) [37], the peaks with relatively lower $E_B$ correspond to $Sn^0$ (Sn metal) components. The peak intensity of the $Sn^0$ relative to that of the oxidized $Sn^{2+}$ (SnO) of the irradiated area is lower than that of the as-grown one, indicating that the oxidized Sn layer in the irradiated area is thicker than that of the as-grown one. This result can be attributed to the physical damage caused by FIB irradiation as shown in Fig. 3(b), which resulted in an increase in defects and the formation of a deeper oxidized layer.

The $Sn^0$ $3d_{5/2}$ peak positions in the as-grown film and the irradiated area are located at $E_B \sim$ 485.30 eV and 484.96 eV, respectively. The difference between the peak positions indicates that the chemical state of the $Sn^0$ is changed by the irradiation of FIB. Because the difference between the peak positions of the $Sn^0$ (0.45 eV) is nearly the same as that between α-Sn and β-Sn [38-40], it is expected that α-Sn is changed to β-Sn by the FIB irradiation.

To elucidate the electronic structure near the $E_F$ of the as-grown and irradiated Sn parts, the valence-band (VB) spectra of the samples are compared. Figure 3(c) shows the VB spectra of these samples. These VB spectra are nearly identical except for the spectral features near $E_F$. Figure 3(d) indicates the VB spectrum of the irradiated area has a finite intensity at $E_F$ with the step of the Fermi-Dirac distribution function, called Fermi edge. This result indicates that the irradiated area is metallic. In contrast, the intensity of the VB spectrum of the as-grown sample gradually decreases with approaching $E_F$. This spectral feature is typical of the density of states of semi-metallic compounds and consistent with the calculated results for semi-metallic α-Sn [41]. This result is consistent with the TDS nature of the as-grown α-Sn/InSb film. Based on the $Sn^0$ peak position and the observation of the Fermi edge, it is strongly suggested that the chemical state of Sn in the irradiated area is superconducting β-Sn. Considering the side effects of FIB and the fact that α-Sn is stable only at low temperatures, the local annealing induced by the FIB process transforms α-Sn to β-Sn.

As to the effect of the local annealing, spreading heat by FIB is predicted to distribute in tens of nm from the outside of the irradiated area [32]. To elucidate the spatial distribution and the chemical states of the α-Sn/β-Sn interface experimentally, we measured XPS spectra near the interface between the as-grown and the irradiated area of the α-Sn thin film. Figure 4(a) shows the Sn $3d_{5/2}$ core-level spectra of the α-Sn, the irradiated area, and the interface region, respectively. Here, the spectra are normalized to the height of the $Sn^{2+}$. While the spectral line shapes of the oxidized $Sn^{2+}$ components are nearly identical among the spectra, the peak positions of the $Sn^0$ components seem to be different. To analyze the interface state, the intrinsic $Sn^0$ components are obtained by subtracting the extrinsic $Sn^{2+}$ components from the Sn $3d_{5/2}$ spectra, where the peaks of the $Sn^{2+}$ components are fitted by Voigt functions. Figure 4(b) shows the $Sn^0$ peak of the interface fitted with a linear combination of the α-Sn and β-Sn contributions to the $Sn^0$ component. The fitting well reproduces the $Sn^0$ peak of the interface, indicating that the chemical states of the interface mainly consist of α-Sn and β-Sn, not a specific state near the interface. Considering the spot size of the incident X-ray beam of some hundred mm, the result suggests that the spreading area of the local annealing is less than a hundred mm scale, consistent with the simulation result [34]. The results suggest that FIB treatment enables us to create nm-order patterns of superconducting β-Sn in a TDS α-Sn thin film, which is essential for fabricating superconductor/TDS devices based on α-Sn.

Finally, we discuss the quantitative difference of the binding energy ($\Delta E_B$) between α-Sn and β-Sn. The difference in the peak positions reflects the different chemical states. In general, $E_B$ is represented as

$$E_B = KQ + E_0 + V - E_R, \quad (1)$$

where $K$ is the empirical constant, $Q$ is the charge number, $E_0$ is the $E_B$ of a neutral free atom, $V$ is the potential energy produced by all the other surrounding atoms, and $E_R$ is the relaxation energy that reflects a decay process of a core-hole screened by intra- and inter-valence electrons. In general, $\Delta E_B$ between two compounds is represented as

$$\Delta E_B = K\Delta Q + \Delta V - \Delta E_R \quad (2)$$

In the case of $\Delta E_B$ for allotropes, the first term in Eq. (2) is equal to 0 because the charge number is common in allotropes. The second term of D$V$ for allotropes comes from the difference in the crystal structure. Since the nearest neighbor atoms likely affect $V$, a coordination number $z$ is expected to be a dominant factor for $\Delta V$ [39]. The third term of D$E_R$ reflects the difference in the metallicity. For carbon allotropes with similar metallicity ($\Delta E_R \sim 0$), $\Delta E_B$ ($\sim \Delta V$) is well explained by the difference of $z$ based on bond-order-length-strength (BOLS) theory [42]. In this theory, it is assumed that if a chemical bond breaks, the remaining bonds become shorter and stronger [43,44]. According to BOLS theory, the ratio of $E_B$ between different coordination numbers ($z$ and $z'$) is given by

$$\frac{E_B(z') - E_0}{E_B(z) - E_0} = \frac{C(z')^{-m}}{C(z)^{-m}},$$

(3)

where $C$ is the bond contraction coefficient and $m$ is the bond nature indicator. Here, $m$ is about 1 [45] and $E_0$ is 479.60 eV [46] for Sn. $C(z)$ is represented by the following equation,

(4)

$$C(z) = \frac{2}{1 + \exp\left(\frac{12-z}{8z}\right)}.$$

As shown in Fig. 5(a), α-Sn has a diamond structure with $z = 4$, where the interatomic Sn-Sn distance from the four nearest neighbor atoms is 2.29 Å. On the other hand, in the β-Sn having a tetragonal structure, the coordination number $z$ is approximately 6 because the second-nearest interatomic Sn-Sn distance of 3.18 Å with two atoms is nearly the same as the nearest neighbor atomic distance of 3.02 Å with four atoms. Using these coordination numbers and the Sn $3d_{5/2}$ peak position of β-Sn, the value of $\Delta E_B$ is calculated to be 0.38 eV from Eqs. (3) and (4). This value is close to the experimental $\Delta E_B$ of 0.34 eV. The remaining difference of $\Delta E_B$ between the calculated and experimental values (~0.04 eV) likely comes from $\Delta E_R$ in Eq. (2) reflecting the metallicity. This result indicates that the chemical states of Sn allotropes are predominantly affected by the change of the crystal structure, i.e., the coordination number $z$.

**Conclusion**

In summary, we have performed transport, TEM, and XPS measurements on a FIB-irradiated α-Sn thin film and clarified the allotropic transition from Dirac semimetal α-Sn to superconductor β-Sn by FIB irradiation. The transport measurements demonstrate that the FIB-irradiated area becomes superconductor with its transition temperature of ~ 4 K. The TEM observations indicate that the irradiated area remains crystallized even with physical damages by FIB and the boundary regions of the different phases are less than several tens nm scale. The XPS results suggest that TDS α-Sn transforms into superconducting β-Sn when being irradiated with FIB. Based on the findings, it is likely that the local annealing, one of the side effects of FIB, induces the allotrope transformation from α-Sn to β-Sn. From the quantitative analysis of the chemical states, we have found that the changes in the coordination numbers accompanied by the allotrope transition primality affect the chemical states rather than the metallicity. Our results provide an insight into the phase transition of Sn upon FIB irradiation and suggest a new method for fabricating devices of TDS α-Sn combined with superconductor β-Sn by FIB lithography [27].

**Acknowledgement**

This work was supported by Grants-in-Aid for Scientific Research Grants-in-Aid for Scientific Research (19K21961, 20H05650, 23K17324), CREST (JPMJCR1777), PRESTO (JPMJPR19LB), ERATO (JPMJER2202) of Japan Science and Technology Agency, the UTEC-UTokyo FSI research granting program, the Murata Science Foundation. This work was partially supported by the Spintronics Research Network of Japan (Spin-RNJ). This work was performed under the Shared Use Program of Japan Atomic Energy Agency (JAEA) Facilities (Proposal No. 2022A-E23) supported by JAEA Advanced Characterization Nanotechnology Platform as a program of "Nanotechnology Platform" of the Ministry of Education, Culture, Sports, Science and Technology (MEXT) (Proposals No. JPMXP1222AE0031). A part of this work was supported by the "Advanced Research Infrastructure for Materials and Nanotechnology in Japan (ARIM)" of the MEXT (Proposals No. JPMXP1223UT1014). K. Ishihara acknowledges support

Figure 1

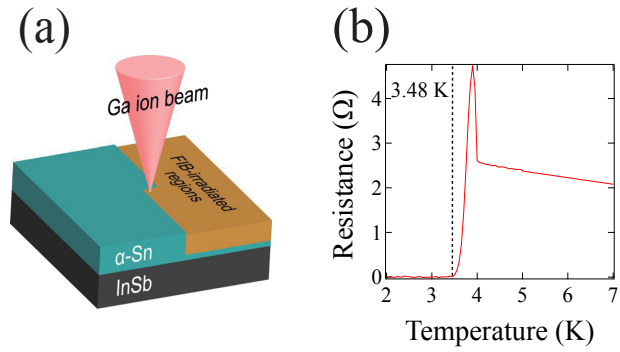

FIG .1. Schematic image and transport properties of the FIB-irradiated α-Sn thin film. (a) The sample is prepared by irradiating FIB on half of the α-Sn thin film area. We used a Ga ion beam with an acceleration voltage of 30 kV and a current value of 7.87 pA. (b) Temperature dependence of the electrical resistance of the FIB-irradiated Sn area. We confirmed that the superconducting transition occurs below 4 K.

Figure 2

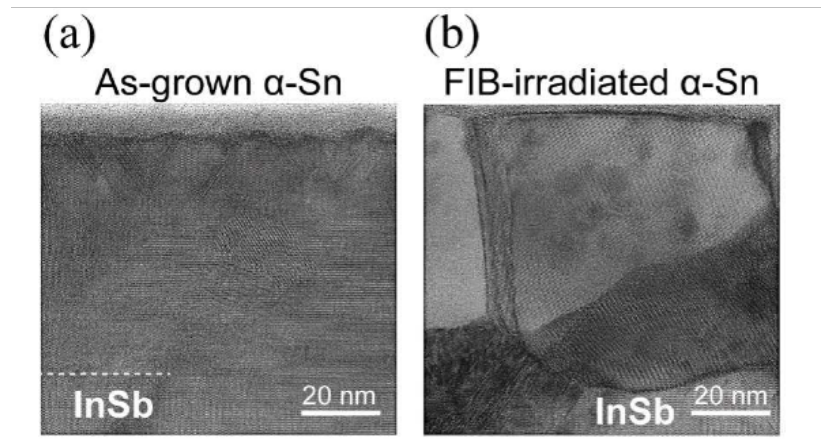

FIG. 2. Cross-sectional transmission electron microscopy (TEM) images of the as-grown and the FIB-irradiated α-Sn area on an InSb (001) substrate. (a) Cross-sectional TEM lattice image of the as-grown α-Sn area. As-grown α-Sn shows a diamond-type crystal structure in the entire film. (b) Cross-sectional TEM lattice image of the FIB-irradiated α-Sn area. The morphology is changed by FIB irradiation, and multiple crystal grains are formed.

Figure 3

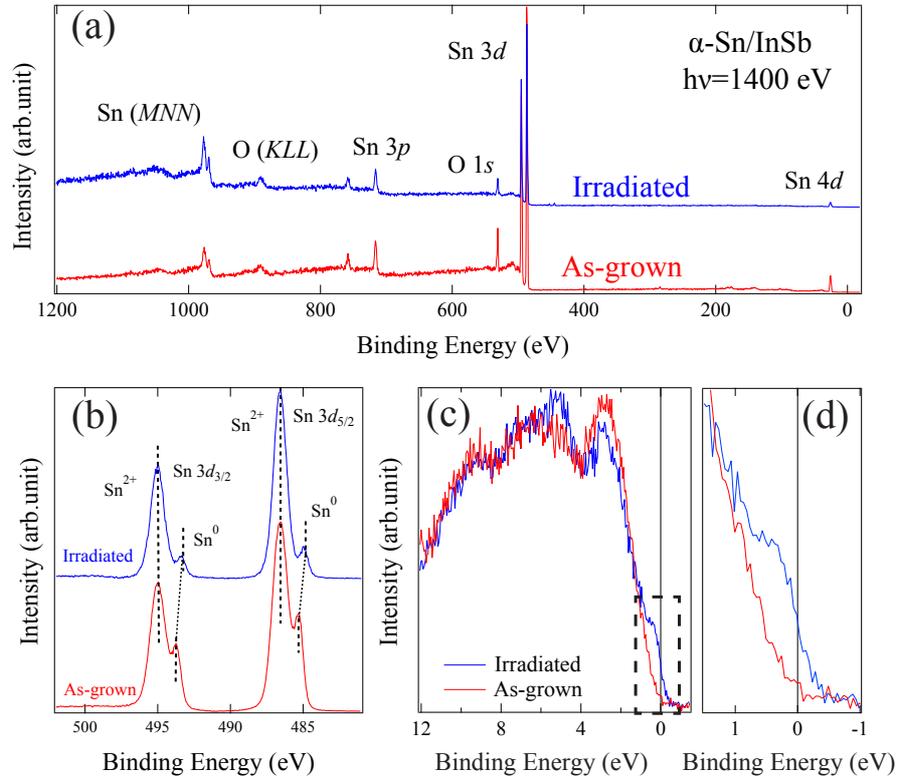

FIG. 3. XPS spectra of the α-Sn/InSb thin film irradiated by FIB. (a) A wide spectrum of the as-grown and the irradiated area. (b) Sn 3$d$ core-level spectra of as-grown and the irradiated area. (c),(d) Valence band spectra of the as-grown film and the irradiated area, which are shown in a wide VB range and enlarged plot near $E_F$. $E_F$ is 0 eV in the figure. The as-grown spectrum reaches zero at the zero binding energy. The irradiated area spectrum has a sharp step near the zero binding energy. All the spectra have been taken with $h\nu$ = 1400 eV.

Figure 4

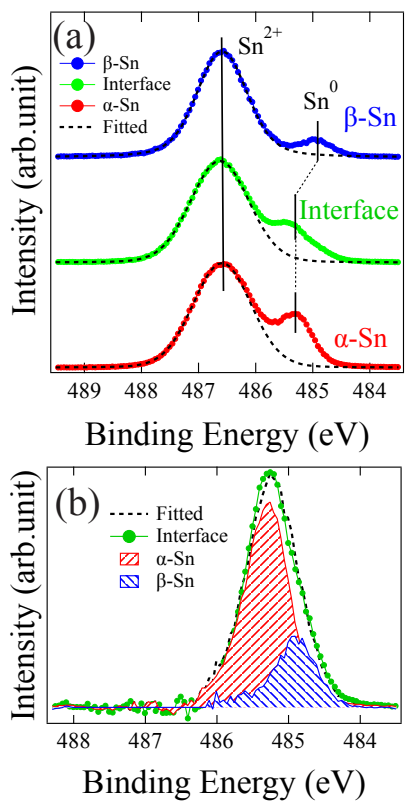

FIG. 4. Sn $3d$ core-level spectra of the as-grown thin film, the irradiated area, and the interface between them. (a) Peak fitting for the oxide Sn peak for Sn $3d_{5/2}$ core-level spectra. (b) $Sn^0$ component of the interface. The spectrum is fitted by a linear combination of α-Sn and β-Sn contributions.

Figure 5

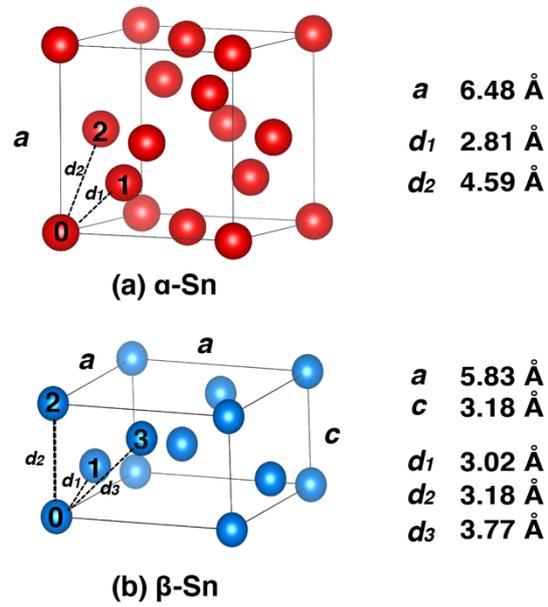

FIG. 5. Crystal structures and intraatomic distances of the Sn allotropes; (a) α-Sn. (b) β-Sn. The numbers on the atoms indicate the distance order from the atom with zero. The structure is drawn by VESTA 3 [47].